\documentstyle{mn}

\newif\ifAMStwofonts

\ifoldfss
  \ifCUPmtlplainloaded \else
   \NewTextAlphabet{textbfit} {cmbxti10} {}
    \NewTextAlphabet{textbfss} {cmssbx10} {}
    \NewMathAlphabet{mathbfit} {cmbxti10} {} 
    \NewMathAlphabet{mathbfss} {cmssbx10} {} 
  \fi
  \ifAMStwofonts
    \ifCUPmtlplainloaded \else
      \NewSymbolFont{upmath} {eurm10}
      \NewSymbolFont{AMSa} {msam10}
      \NewMathSymbol{\upi}     {0}{upmath}{19}
      \NewMathSymbol{\umu}     {0}{upmath}{16}
      \NewMathSymbol{\upartial}{0}{upmath}{40}
      \NewMathSymbol{\leqslant}{3}{AMSa}{36}
      \NewMathSymbol{\geqslant}{3}{AMSa}{3E}

    \fi
  \fi
\fi 

\ifnfssone
  \newmathalphabet{\mathit}
  \addtoversion{normal}{\mathit}{cmr}{m}{it}
  \addtoversion{bold}{\mathit}{cmr}{bx}{it}
  \newmathalphabet{\mathbfit} 
  \addtoversion{normal}{\mathbfit}{cmr}{bx}{it}
  \addtoversion{bold}{\mathbfit}{cmr}{bx}{it}
  \newmathalphabet{\mathbfss} 
  \addtoversion{normal}{\mathbfss}{cmss}{bx}{n}
  \addtoversion{bold}{\mathbfss}{cmss}{bx}{n}
  \ifAMStwofonts
    \ifCUPmtlplainloaded \else
      \UseAMStwoboldmath
      \makeatletter
      \new@mathgroup\upmath@group
      \define@mathgroup\mv@normal\upmath@group{eur}{m}{n}
      \define@mathgroup\mv@bold\upmath@group{eur}{b}{n}
      \edef\UPM{\hexnumber\upmath@group}
      \new@mathgroup\amsa@group
      \define@mathgroup\mv@normal\amsa@group{msa}{m}{n}
      \define@mathgroup\mv@bold\amsa@group{msa}{m}{n}
      \edef\AMSa{\hexnumber\amsa@group}
      \makeatother
      \mathchardef\upi="0\UPM19
      \mathchardef\umu="0\UPM16
      \mathchardef\upartial="0\UPM40
      \mathchardef\leqslant="3\AMSa36
      \mathchardef\geqslant="3\AMSa3E
    \fi
  \fi
\fi 

\ifnfsstwo
  \DeclareMathAlphabet{\mathbfit}{OT1}{cmr}{bx}{it}
  \SetMathAlphabet\mathbfit{bold}{OT1}{cmr}{bx}{it}
  \DeclareMathAlphabet{\mathbfss}{OT1}{cmss}{bx}{n}
  \SetMathAlphabet\mathbfss{bold}{OT1}{cmss}{bx}{n}
  \ifAMStwofonts
    \ifCUPmtlplainloaded \else
      \DeclareSymbolFont{UPM}{U}{eur}{m}{n}
      \SetSymbolFont{UPM}{bold}{U}{eur}{b}{n}
      \DeclareSymbolFont{AMSa}{U}{msa}{m}{n}
      \DeclareMathSymbol{\upi}{0}{UPM}{"19}
      \DeclareMathSymbol{\umu}{0}{UPM}{"16}
      \DeclareMathSymbol{\upartial}{0}{UPM}{"40}
      \DeclareMathSymbol{\leqslant}{3}{AMSa}{"36}
      \DeclareMathSymbol{\geqslant}{3}{AMSa}{"3E}
    \fi
  \fi
\fi 

\ifCUPmtlplainloaded \else
  \ifAMStwofonts \else 
    \def\upi{\pi}
    \def\umu{\mu}
    \def\upartial{\partial}
  \fi
\fi

\title{A Phenomenological Model for QSOs}
\author[S. Nasiri and F. Tabatabaei]
       {S. Nasiri$^{1,2}$
       \& F.Tabatabaei$^{1,3}$
       \\ $^1$Institute for Advanced Studies in Basic Sciences, Zanjan, Iran\\
       $^2$Department of Physics, Zanjan University, Zanjan,
       Iran\\
       $^3$Department of Physics, Damghan University of Basic Sciences,
       Damghan.}
\date{July 2000}

\pagerange{\pageref{firstpage}--\pageref{lastpage}} \pubyear{}

\begin{document}

\maketitle

\label{firstpage}

\begin{abstract}
A combined model on the basis of generalized Field­Colgate and
Terlevich­ Melnick models are proposed for QSOs. Using LBQS data,
it seems that the predictions of the model is confirmed by
observations. The behavior of comoving density versus redshift is
consistent with LDDE for QSQs. Considering the cosmic evolution of
SFR, a unified aspect for origin and evolution of QSOs and
ordinary galaxies is implied by this model.
\end{abstract}


\section{INTRODUCTION}
\label{sect:intro}  

Since their discovery, different models have been proposed for
explaining the observational properties, the origin and the
evolution of QSOs. Among them, the remarkable models are: the
standard model, Terlevich ­ Melnick (TM) model[1,2,3] and Field ­
Colgate (FC) model [4]. Besides the difficulties existing with
each model, a considerable knowledge about these objects is
established upon them. These models emphasize on two main points:
a) the origin and evolution of QSOs and b) their inherent
properties. Here we combine a generalized version of the FC model
[5] called GFC model with the TM model. Both of these are
starburst models and have some conceptual features in common. The
difficulties that an individual model are faced on, is removed for
combined model. The by products of the resultant model are: a) the
decay mechanism which is shown to be supported by observations
using LBQS data[6] the study of comoving density versus redshift
satisfies the luminosity dependent density evolution (LDDE) for
QSOs, c) a unified aspect for the origin and the evolution of QSOs
and the ordinary galaxies is another implication of the combined
model considering the cosmic evolution of the star formation rate
(SFR) for these objects. In sec.2 we give a brief review of the
existing models for QSOs. In sec.3 the combined model is described
and sec.4 is ordered to observational investigation of evolution
of QSOs and their correlation with galaxies. In sec.5 the
concluding remarks are given.

\section{A BRIEF REVIEW OF EXISTING MODELS}

The early days of AGN research, produced a wide variety of
remarkable models that were designed to explain the enormous
amount of energy output of QSOs. However most of this models have
failed in one way or another to explain for various QSO
characteristics. Here we present a brief review of the most
important models such as standard model, TM model and FC model.
Most researchers agree that the energy source in QSOs is primarily
gravitational and involves large concentrations of matter such as
super massive stars or massive black holes[7,8,9]. These has
shifted the aim of theoretical and observational works towards the
study of the properties of such objects and their environments.
They have designed a model in base of accretion discs circling
massive black holes. In this model mass of the central black hole
for a QSO with typical luminosity, $10^{46} ergs^{-1}$ is assumed
to be about $10^{8} m_{\odot}$. They also postulate the existence
of small high density clouds very close to the nucleolus where
only broad permitted lines (BLR) are formed and a more extended
system of low density filaments where narrow forbidden and
permitted lines are formed (NLR). The line widths of the BLR (up
to 10000 $Kms^{-1}$ FWHM) are assumed to reflect the motions of
the ensemble of cold ($T \sim 10^{4} K$) and dense ($n \sim
10^{10} cm^{-3}$)clouds moving in the gravitational field of the
massive central object. The ionizing spectrum is assumed to follow
a power law of the form $f_{\nu} = \nu^{-1.0}$ up to a few hundred
KeV. Other typical parameters of these clouds, also derived from
the photoionization models [10](e.g. Collin-Souffrin 1990), are
the ionization parameter $U\sim 2\times 10^{-3}$ and column
density $\sigma\sim 10^{23}cm^{-2}$. While the black hole model
for the QSO central engine is successful in accounting for many of
the observed properties of the BLR of QSO, it remains
unsatisfactory in that it requires a large number of arbitrary
parameters which can not be predicted from theory and are freely
adjusted to match the observations [11]. Meanwhile existing the
accretion disk, the ubiquitous component of this model has not
been proved through the polarization studies on both sides of the
Lyman limit for intermediate-redshift QSOs observed by HST[12].

 Another model which has not been
definitively discredited is the model proposed by Terlevich and
collaborators. The model is based on the nuclear starburst
scenario. They assume that the observed activity of QSOs is the
direct consequence of the evolution of a massive young cluster of
coeval stars in the high metal abundance HII regions of early type
galaxies [13]. In the other words, QSOs are postulated to be the
evolutionary phase of elliptical galaxies. During this phase, most
of the bolometric luminosity is emitted by the young stars, while
the broad permitted emission lines and their variability are
mainly due to rapidly evolving compact supernova remnants
(SNR)[14]. Theoretical computations of the evolution of SNR in
dense molecular clouds show that after sweeping up a small amount
of gas and when their sizes are only few light weeks across, these
remnants become strongly radiative. They deposit most of their
kinetic energy in very short time­scales, thus reaching very high
luminosity. Because of the large shock velocities, most of the
energy is radiated in the extreme UV and X­ray region of the
spectrum [15]. Terlevich \emph{et al} studied the evolution of SNR
in a high­density medium ($n\sim 10^ 7 cm^{-3}$) the observed
values of parameters of the BLR with density of the medium as the
free parameter. But some problems exist with this starburst model.
The rapid variability of the X­ray radiation is not derived from
this model. Further, the size of a typical QSO is assumed to be
equal to the size of the core of a typical elliptical galaxy that
have a radius of about 200Pc [13]. But imaging studies with HST
show that AGNs remain unresolved at the highest currently
attainable spatial resolution 0.05''[11].

 Other astronomers not only do not consider the QSOs
as an evolutionary phase of early­type galaxies, but also consider
them as an extreme case of galaxies with independent evolutions.
According to this idea, Field and Colgate propounded a starburst
model, in which, galaxies and QSOs were different aspects of a
same phenomenon. One generally believes that the galaxies, as
separate units, are originated through some sort of gravitational
instability. Moreover a fluctuation in density either developed or
pre­existed in the proto­galaxies from which the galaxies were
start to form. As a fluctuation grew in mass, it collapsed under
the action of gravity, cooled and eventually a galaxy was formed.
If one accepts that QSOs are extreme case of galaxies, one must
seek for some characteristic physical parameters which are
responsible for the observational differences of these objects
with the galaxies. Field and Colgate [4] considered the angular
velocity of proto­galaxies as a characteristic parameter. The FC
model assumes that the size of the galaxies, average mass of their
constituent stars and their total energy output depend on the
rotation rate of the proto­galaxies or equally on the balance of
the gravity with the centrifugal force at the end of the
collapsing process. As an example assume two proto­galaxies with
the same initial masses and sizes, but one with an angular
velocity ten times that of the other. The centrifugal force will
then be hundred times weaker for the slowly rotating proto­galaxy.
Such an object will eventually be about hundred times smaller in
size and have, on the average, stars of about fifty times more
massive than that of the fast rotating one. Assuming that their
constituent stars are of main sequence type, the compact object
will generate about $2.5\times 10^{3} $ times more energy than the
extended one. According to the FC model, the compact and extended
sources in the preceding example are the representatives for a QSO
and an ordinary galaxy, respectively. While the FC model as a
starburst model was able to describe energy problem and size of
the QSOs, it could not explain most of their observed properties
such as variability, their radio emission, unresolved images, the
forbidden emission lines in the spectra and their intensity
ratios. However, as mentioned before, the TM model, in spite of
the fact that it could not satisfy the observed sizes of QSOs, was
able to explains the other properties, satisfactorily. A claim
which may arise here is the possibility of combining the FC and TM
models as two starburst models, to reduce their individual
problems. The authors recently have generalized the FC model by
assuming the specific angular momentum (SAM) as the characteristic
parameter for proto­galaxies [5]. This quantity which mainly
excludes the effect of the initial mass of the proto­galaxies is
assumed to be constant during the contraction procedure and is
replaced by the angular velocity in the FC model. They called it
generalized FC or GFC model and showed that the expected relation
between the SAM and specific luminosity (SL) predicted by GFC
model is satisfied by observations. They used the LEDA database
for galaxies with different morphological types and showed as SL
increases the SAM decreases. A by product of GFC model is the so
called ''decay mechanism'' for QSOs. If we assumes that the QSOs
are evolved from the slowly rotating proto­galaxies, and
therefore, posses much massive stars, one should accept that they
must evolve faster than the ordinary galaxies as well. Thus, the
QSOs formed in this way would have a half life proportional to the
inverse square of their masses if presumably populated by the main
sequence type of stars. They will evolve about $10^ 4$ times
faster than the correspondingly ordinary galaxies. The QSOs
populated, on the average, by stars with the masses greater than
eight solar mass, may eventually disappear from the contact with
the rest of the universe as a result of collapse after consuming
their energy sources. This process, if done, may lead to an
evolutionary decay mechanism working for these objects in the
course of time. We will examine this phenomenon in Sec.3 using
observational data. In addition, using the same database, the
plots of the number of compact (C) and the diffuse (D) galaxies
versus their absolute magnitude shows that, as the object gets to
be more and more compact becomes more and more luminous[5]. As
outlined before, the dynamical aspects of the FC model is not
capable of explaining some other observational properties of QSOs.
The same weakness already exist for GFC. To remedy this problem
will propose another model on the basis of GFC and TM
considerations in the next section.


\subsection{THE COMBINED MODEL}
\label{sect:title}

One may use the common features of the GFC and TM models, to
obtain a new starburst model which includes the advantages and
excludes the disadvantages of the previous models. In this
respect, the following points are noteworthy:\\ a) In the TM
model, QSOs are considered as an evolutionary phase of elliptical
galaxies, whereas, in the GFC model galaxies and QSOs are
different manifestation of the same phenomenon that evolve
independently biased by different initial conditions. \\ b)Either
GFC or TM model, assumes the massive stars as the source of the
bolometric luminosity of QSOs. However, the BLR and NLR in the
spectrum of the QSOs, their intensity ratios and variability could
not be interpreted by GFC model. While in the TM model, thier
observational properties are explained by assuming SNR interacting
with a relatively dense interstellar medium. \\ c) Although the TM
model is successful in accounting for many of the observed
properties of the BLR of QSOs, with only one free parameter, it is
not capable of describing the small size of these objects. This is
because of the considerations of QSOs as the core of elliptical
galaxies with typical size$\sim 200pc$ (about hundred times more
than observed values), in an evolutionary stage of the same
galaxies. Whereas in the GFC model, the size difficulty does not
exist. Here one can adjust the initial SAM of proto­galaxies to
arrive finally with the observed size of QSOs.

Therefore, it is seen that except for the initial considerations
not only there is not a discrepancy between this two models, but
also they complement each other. So it seems that a combined form
of them may be more efficient model in explaining the observed
properties of QSOs. Thus, one can assume, as mentioned before, the
QSOs are eventually made from the proto­galaxies with relatively
low initial SAM compared with that of the proto­galaxies which
evolve finally to ordinary galaxies. Note that in the combined
model two independent parameters, i.e. the SAM assumed by GFC
model and the interstellar density conveniently chosen by TM
model, determine the final state of the proto­galaxy.
\section{THE COMBINED MODEL AND PHENOMENOLOGICAL RESULTS }
The observational implications of the combined model may be
examined using the available complete surveys. Here we have used
the Large Bright QSO survey (LBQS)[6] containing 1055 QSOs in the
magnitude range of $16.0< m_{B_{j}} <18.85$ and redshift range of
$0.2< z <3.4$. The proposed combined model may
have the following predictions,\\
 1. Decay mechanism,\\
 2. Galaxy­QSO\\
unification. In the following subsections we explain the above
results separately.

\subsection{ DECAY MECHANISM}
According to the combined model the QSOs must evolve faster than
the ordinary galaxies. This is due to the high rate of energy
production of their constituent massive stars. Therefore, one
expects that the nearby QSOs must disappear from the contact of
the rest of the universe by the known massive stars evolution
scenario. Moreover, the disappearance must be most pronounced for
luminous QSOs. To check the above prediction one possible way is
to consider the space distribution of QSOs. By considering the
look back time effect on relatively distance objects such as QSOs,
the above arguments lead to a nonuniform space distribution for
them. In the other words, the plot of comoving number density
versus redshift should reveal relatively more QSOs at far
distances (i.e. at very long times ago). This is not satisfied by
fig.1, that shows more or less uniform distribution for collective
data of LBQS. To obtain the comoving density, we assumed
$H_{0}=75Kms^{-1}Mpc^{-1}$ and $q_{0}=0.0$. However the
discrepancy maybe removed by dividing the full range of magnitudes
of a complete sample of QSOs into different luminosity groups. To
do this, we have classified the data into 4 luminosity groups,
where the absolute magnitude increases with increasing the order
of groups. The group ranges are: (­25.5, ­23.0), (­27.0, ­25.5),
(­28.5, ­27.0), (­30,­28.5). The corresponding distributions are
plotted in figs. 2 to 5. As an overall view, it is clear from
these figures that the decay mechanism exists and is more
pronounced for the luminous quasars. The comoving density of QSOs
associated with the 1st group, i.e. the dimmer QSOs are plotted in
fig.2. They belong to redshift range of $0.2 < z < 0.93$. It shows
that, when we go to the redshifts higher than z = 0.2, the
comoving density decreases so that after $z \sim 0.93$, no QSOs
with this luminosity is seen. This may be due to seeing
restrictions of these dimmer group of QSOs. Also it is seen that
the decay rate goes rather slowly for this group. The situation is
different for 2nd group (fig.3). The comoving density is
distributed in the redshift range of $0.4 < z < 1.53$, and nothing
is observed in the redshift range of $0.2 < z < 0.4$. The lack of
observation of QSOs for 3rd group corresponding to $0.72 < z <
2.53$ is enhanced and nothing is seen for $z < 0.72$. For the most
luminous QSOs, i.e., the 4th group, decay mechanism has been going
more rapidly. So that no QSOs is observed for $z < 1.08$. Note
that, decreasing the comoving density of QSOs with increasing the
group order for lower redshifts is not due to the instrumental and
seeing limitations, because these kind of limitations become
unimportant at high luminosities. The results obtained above, may
be interpreted from the point of view of luminosity function of
QSOs. For each group one can assume a distinct pure density
evolution (PDE). However, the behavior of PDE is considerably
different for different groups. Thus, for the collective LBQS data
the luminosity dependent density evolution (LDDE) is more
convenient.

\begin{figure}
 \vbox to2.5in{\rule{0pt}{2.5in}}
\includegraphics{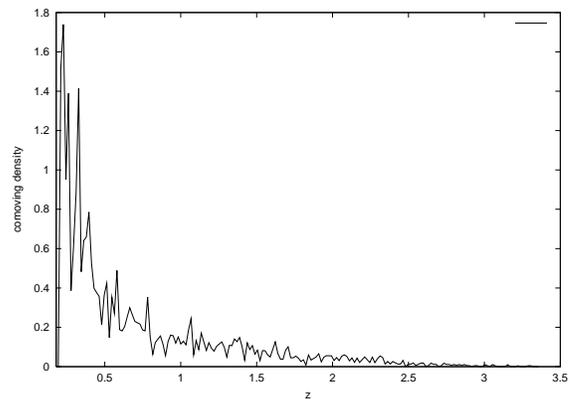} \caption{ The behavior of comoving number
density (in units of $Gpc^{-3}$) versus redshift.}
\end{figure}

\begin{figure}
 \vbox to2.5in{\rule{0pt}{2.5in}}
\includegraphics{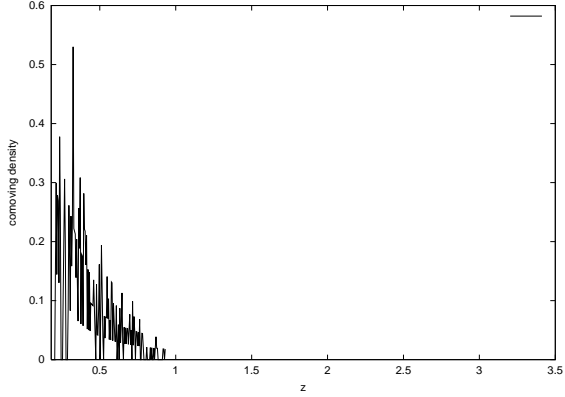} \caption{ The same as Fig. 1 for the first
luminosity group.}
\end{figure}

\begin{figure}
 \vbox to2.5in{\rule{0pt}{2.5in}}
\includegraphics{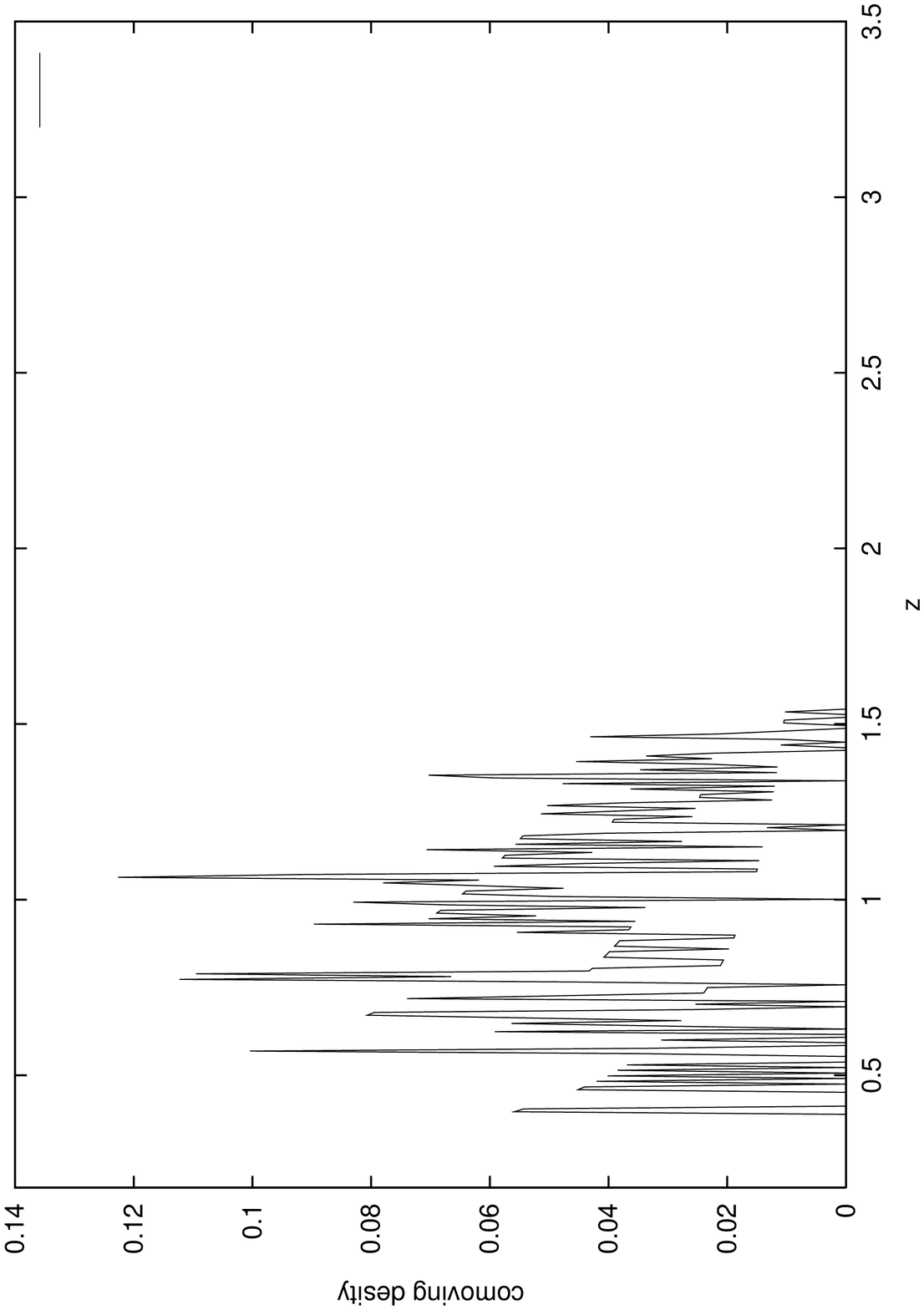} \caption{ The same as Fig. 1 for the 2nd
luminosity group.}
\end{figure}

\begin{figure}
 \vbox to2.5in{\rule{0pt}{2.5in}}
\includegraphics{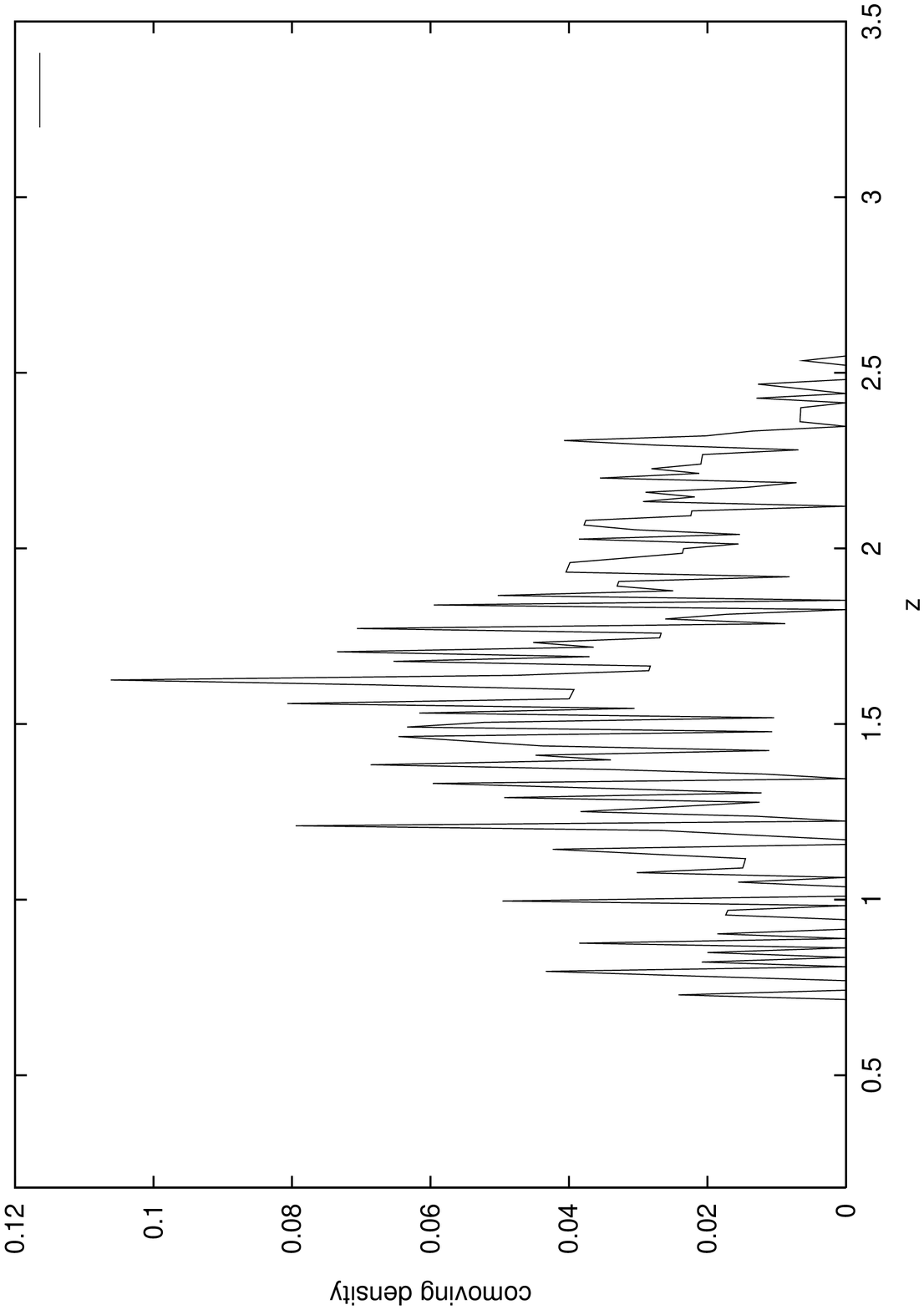} \caption{ The same as Fig. 1 for the 3rd
luminosity group.}
\end{figure}

\begin{figure}
 \vbox to2.5in{\rule{0pt}{2.5in}}
\includegraphics{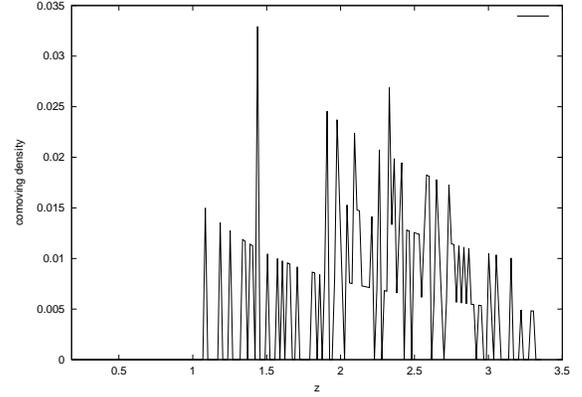} \caption{ The same as Fig. 1 for the 4th
luminosity group.}
\end{figure}

\subsection{QSOs AS EXTREME CASE OF GALAXIES}
It is shown by Boyle and Terlevich that the cosmological evolution
of star formation rate (SFR) of galaxies are almost similar to the
luminosity density evolution of QSOs[16]. Franceschini \emph{et
al}, recently have shown that the similarity between luminosity
density of QSOs and SFR in elliptical galaxies are more
remarkable[17]. Thus, it seems that there exist a relation between
SFR and evolution of QSOs, indicating that a nuclear starburst is
powering much of the QSOs luminosity as assumed in combined model.
By the above arguments, one expects the evolution rate decreases
from QSOs to elliptical galaxies and then to spirals. In the other
words, QSOs evolve rapidly compared with the galaxies and among
the galaxies, the ellipticals evolve faster than the spirals[5].
\subsection{CONCLUDING REMARKS}

The generalized form of FC model is combined with TM model to
obtain a combined model for QSOs. The resultant model is capable
of explaining the most observational properties of these objects
such as energy production, size, BLR, NLR, variability, etc. The
model assumes QSOs as evolved form of slowly rotating
proto­galaxies. A decay mechanism, acquired by this assumptions,
is supported by LBQS data classified by different luminosity
groups. On the other hand the ordinary galaxies are assumed to
evolve from relatively rapidly rotating galaxies, which is a
motivation for considering the QSOs as extreme case of galaxies.
Thus in the framework of the combined model QSOs and the galaxies
of different morphologies are considered to have the same origin
with different initial conditions which affect their evolution
rate at the later times. The investigation of behavior of the
comoving density in terms of redshift for different luminosity
groups supports the luminosity dependent density evolution for
QSOs.


\bsp

\label{lastpage}

\end{document}